# Spin Relaxation in Germanium Nanowires

Ashish Kumar[a)] and Bahniman Ghosh

*Department of Electrical Engineering, Indian Institute of Technology, Kanpur 208016, India*

Abstract-We use semiclassical Monte Carlo approach along with spin density matrix calculations to model spin polarized electron transport. The model is applied to germanium nanowires and germanium two dimensional channels to study and compare spin relaxation between them. Spin dephasing in germanium occurs because of Rashba Spin Orbit Interaction (structural inversion asymmetry) which gives rise to D'yakonov-Perel(DP) relaxation. In germanium spin flip scattering due to Elliot-Yafet(EY) mechanism also leads to spin relaxation.

The spin relaxation tests for both 1-D and 2-D channels are carried out at different values of temperature and driving electric field and the variation in spin relaxation length is recorded. Spin relaxation length in a nanowire is found to be much higher than that in a 2-D channel due to suppression of DP relaxation in a nanowire. At lower temperatures the spin relaxation length increases. This suggests that spin relaxation in germanium occurs slowly in a 1-D channel (nanowires) and at lower temperatures. The electric field dependence of spin relaxation length was found to be very weak.

Keywords: Spin Relaxation, Monte Carlo, Nanowires, Germanium.

a). Electronic mail : ashish12.kumar@gmail.com

# I. INTRODUCTION

Of late, intensive experimental and theoretical studies have been conducted on the physics of electron spins due to the enormous promise displayed by the spin-based devices[1]. Spin transport in semiconductors has been continuously investigated due to the possibility of integration of spintronics with semiconductor technology. This integration has attracted huge research interest due to its prospects[2-5] in implementing novel devices that can operate at much less power levels and higher processing speeds. Spintronics based devices thus promise highly improved performance over their contemporary electronic counterparts. The suitability of present semiconductor materials in spintronics based applications needs to be established to bring about successful integration of the two and to be able to achieve the advantages listed above.

The basic idea of the spintronic based devices is to use the spin degree of freedom. At the source information is encoded as spin state of individual electrons and is then injected into the material. During its motion in the material, the electrons undergo scattering and hence the electron spin states relax or depolarize as they move in the channel. This is the process of spin relaxation. Spin detection is done at the drain. Our paper here deals with the second process of spin relaxation in a material. Spin relaxation lengths or spin dephasing lengths represent the distance from the source in which the spin polarization of an ensemble of electrons gets randomized and thus loses the information. We do not want the electrons to lose encoded information before the operation is complete. Thus information regarding spin relaxation lengths is critical to realize any useful spintronic based device.

Intensive theoretical and experimental research has been conducted to study spin relaxation in metals and semiconductors. Several III-V and II-VI materials have been studied to ascertain their spin properties [5-9]. In Ref.[7] spin transport is studied experimentally in GaAs. In Ref. [8] spin polarized transport in GaAs/GaAlAs quantum wells is investigated using Monte Carlo simulations. Spin relaxation in silicon [10,11] has also been studied experimentally. Spin dephasing has also been studied at different conditions of temperature, applied electric field and for different dimensionality of systems [12,13] by researchers in a bid to find the optimum conditions of use of such materials in devices. In Ref.[12] spin transport in GaAs nanowires is modelled at different temperatures and different driving electric fields. Comparison of spin polarized transport in 1D and 2D III-V heterostructures was done in Ref.[13].

Though silicon has been the workhorse of the semiconductor materials since long, germanium has some superior properties [14] to silicon and thus is a material of interest. Germanium has a smaller indirect band gap of 0.66 eV while silicon has an indirect bandgap of 1.12 eV. Germanium has higher electron and hole mobility than silicon. Also germanium has a much lower resistivity than silicon and thus germanium offers great opportunities for device scaling especially when it comes to low drive voltage and high drive current. Recently Ge nanowires have attracted attention due to their possible role in future nanoscale devices such as light emitting diodes (LEDs), logic gates, nanoscale sensors etc. The growing popularity of germanium thus requires us to assess their performance as a spintronics material. However theoretical work on spin relaxation in germanium is still in its very early stages. Much of the work on germanium till now has been experimental [15,16,17]. This growing importance of germanium coupled with the fact that monte carlo simulations have not yet been used to study spin dephasing lengths in germanium (to the best of our knowledge) motivates us to take such a study on germanium.

Spin relaxation in semiconductors can occur via different spin relaxation mechanisms, such as D'yakonov-Perel (DP) [18] mechanism, Bir-Aronov-Pikus [20] mechanism and Elliot-Yafet (EY) [19] mechanism. Bir-Aronov-Pikus mechanism is present in p type semiconductors only and is hence not relevant in our study. In germanium DP mechanism is present. Also being a smaller bandgap material with a high spin orbit coupling (=290meV), EY mechanism is also a dominant spin relaxing mechanism. DP relaxation is a continuous process that occurs even during the free flight time of electrons and it leads to a continuous spin precession about the effective magnetic field. EY relaxation on the other hand is an instantaneous spin-flip scattering and is therefore a discrete process that occurs only during scattering. Therefore EY relaxation is treated as a scattering process.

Semiclassical Monte Carlo approach is used to model electron transport in 2D germanium channels and in 1D germanium nanowires. The Monte Carlo method [21,22,23] along with spin density matrix [23] is used to model spin transport of electrons in both 2-D and 1-D systems. Monte Carlo approach is used since it is able to update spin evolution dynamically in step with the momentum evolution due to electron transport. In conformity with some of the previous works[12,13], improvement in spin relaxation on confinement is observed. Spin relaxation is reinvestigated at different temperatures and different driving electric fields. A similar study was done by us on silicon[24] and this paper is a follow up of the earlier work.

The paper is organized as follows. In Section II we discuss the model used for our simulations. In Section III the parameters used in the simulations are mentioned. The results obtained are shown and a discussion follows on the results in the Section III. The conclusion is presented in the Section IV.

## II. MODEL

A detailed account of the Monte Carlo method [21,22,23] and spin transport model [6,12,23,24] is presented elsewhere. Here we shall be discussing only the key features of the model and the essential modifications from our previous work[24]. The co-ordinate system is so chosen such that $x$ is along the length, $y$ is along the width and $z$ is the along the thickness of the device. In the 2-D system the electrons can move in the $x$ direction and the $y$ direction, while in the 1-D system electrons are free to move only in the $x$ direction.

Germanium is an elemental semiconductor and possesses crystallographic inversion symmetry [15,17]. As a result the Dresselhaus spin orbit interaction is absent [15,17] in germanium. However a transverse electric field breaks the structural inversion symmetry. The structural inversion asymmetry thus present leads to Rashba spin orbit interaction. Rashba spin orbit coupling causes spin relaxation in the channel via D'yakonov-Perel (DP) mechanism.

The conduction band of Ge [25] consists of the four lowest energy <111> L valleys at the edges of the Brilluin zone, the <000> Γ valley at the zone center and the six <100> Δ valleys located near the zone edges. The <111> minimum is 0.14eV below the <000> minimum and 0.18eV below the <100> minimum. We consider here that since the L valleys are lower in energy than the other two valleys, majority of electrons are concentrated in these <111> valleys. Thus we do not consider the other two valleys, i.e. the Γ and the Δ valleys for the sake of our simulation assuming that they are completely depopulated and hence have negligible effects on our final results. Also we assume here that the electric field applied is in the <100> direction and hence the four L valleys remain equivalent.

The scattering processes considered are intravalley and intervalley phonon scattering, surface roughness scattering and ionized impurity scattering. While considering phonon scattering, both optical phonons and acoustic phonons have been taken into account. The electrons will not always remain in the lowest ground subband and will make transitions to the higher subbands. Therefore, subbands are included in the simulation and intervalley and intravalley intersubband scatterings are thus accounted for.

Spin flip scattering [26] via Elliot Yafet mechanism is accounted for in both the the 1-D and 2-D systems. There occurs a finite probability for spin flip due to any perturbing potential even if the perturbation is spin independent (which might be present because of phonons, ionized impurities). The spin relaxation time is given by,

$$\frac{1}{\tau_s^{EY}} = A \left(\frac{k_B T}{E_g}\right)^2 \eta^2 \left(\frac{1-\eta/2}{1-\eta/3}\right)^2 \frac{1}{\tau_p} \tag{8}$$

where $E_g$ is the band gap, $\eta = \Delta/(E_g + \Delta)$ with $\Delta$ as the spin orbit splitting of the valence band. $\tau_p$ is the momentum relaxation time. A is a dimensionless constant and varies from 2 to 6. We have chosen A as 4 for our simulations.

The formula for scattering rates calculations in nanowire and 2-D channels are taken from references [27-33] and have been discussed in our work on silicon[24]

## III. RESULTS AND DISCUSSION

The 2-D channel has 5 nm as the thickness and 125 nm as the width. The nanowire is taken to be of cross-section 5nm x 5nm. The doping density is taken to be 4 x $10^{25}$/m$^3$. The effective field is taken to be 100 kV/cm which is a reasonable value for germanium channels. This effective field acts as the transverse symmetry breaking field and leads to Rashba spin orbit coupling. The Rashba coefficient [34] $\alpha$ is given by,

$$\alpha = \frac{\hbar^2}{2m^*} \frac{\Delta}{E_g} \frac{2E_g+\Delta}{(E_g+\Delta)(3E_g+2\Delta)} eE \tag{9}$$

where $\Delta$ is the spin orbit splitting of the valence band, $e$ is the electronic charge, $m^*$ is the effective mass, $E_g$ is the band gap and $E$ is the transverse electric field. Accounting for the confinement, four subbands are taken for the sake of simulation in each valley for both the channels. The moderate values of driving electric field (100V/cm to 5kV/cm) used ensure that the majority of electrons are restricted to the first four subbands. Also due to very small(5nm) transverse dimensions of the channels, the higher subbands will be very higher up in energy to the effect that we can assume them to be depopulated. Similarly in Ref. [8] 3 subbands were considered for the purpose of determining spin dephasing lengths in a III-V compound. The energy levels of subbands are computed considering an infinite potential well approximation. The other material parameters are considered to be same as that for bulk germanium and are adapted from a standard manual on Monte Carlo simulations [22]. The electrons are injected from the source with initial polarization in the z-direction (along the thickness of the wire). A time step of 0.02 fs was chosen and electrons were run for $1 \times 10^6$ such time steps to ensure that steady state has been reached. Data is recorded for the last 50,000 steps only. The ensemble average is calculated for each component of the spin vector for the last 50,000 steps at each point of the wire.

*a. Spin Relaxation lengths at room temperature(300K) for a driving electric field of 1kV/cm*

A spin transport study was done at room temperature (300K) for both 2-D channels and 1-D nanowires at a moderate driving electric field of 1kV/cm. Since the initial polarization is along the thickness of the wire i.e. in the z-direction, the ensemble averaged x and y components of spin fluctuate around zero (with very small magnitudes) and only the ensemble averaged z component of spin decays along the length. Figure1 shows how the magnitude of spin vector decays along the 2D and 1D channel.

Spin dephasing length in a nanowire is found to be around 210 nm compared to around 12 nm in a 2-D channel. Thus the spin dephasing lengths for a nanowire is about 18 times larger than two dimensional channels. This result bears conformity with similar studies conducted earlier by researchers where improvements in spin relaxation lengths in 1-D channels over 2-D channels have been reported [12,13,35,36].

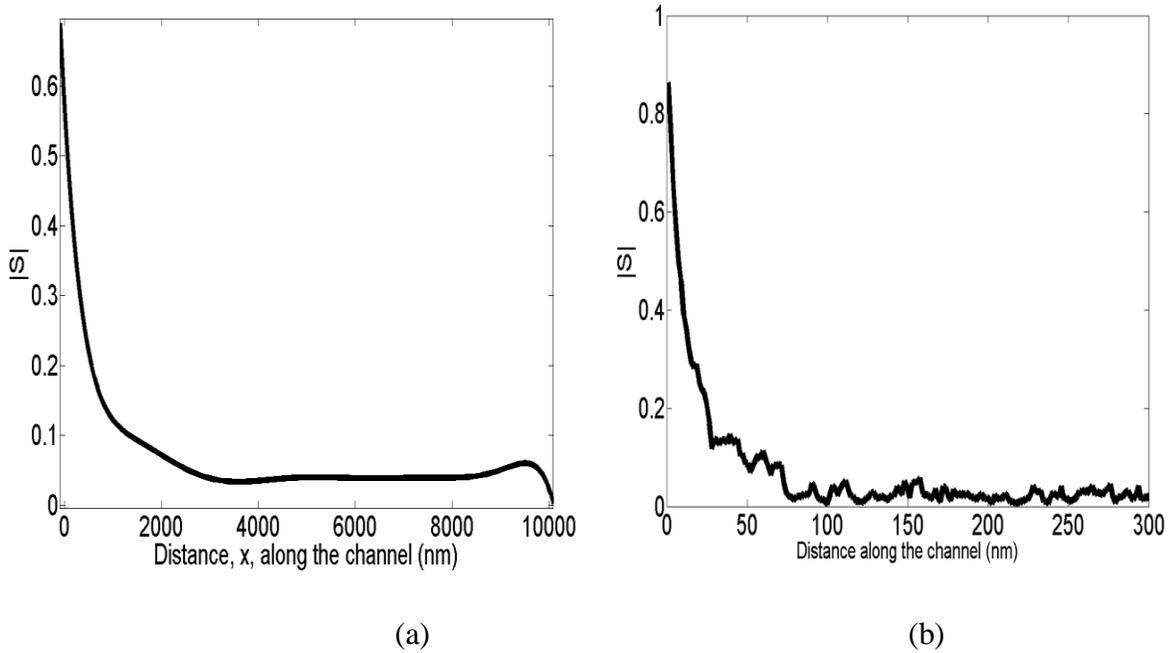

Fig.1: |S| along channel length for electrons in (a) a 1-D channel (b) a 2-D channel.

Explaining this difference in terms of difference in scattering rates in between nanowires and 2-D channels meets with failure owing to the fact that mobility in nanowires has been found to smaller than that in a 2-D channel. The origin of this difference is due to the fact that the dominant spin relaxing mechanism, DP relaxation is suppressed by confinement [37,38]. Thus a nanowire has considerably lesser DP relaxation than a 2-D channel leading to larger dephasing lengths. A detailed explanation to this effect is reported in our previous work [24]. It must be mentioned that our results are consistent with the expectation that spin dephasing increases with the randomness of the motion. In a 2-D channel, randomization of the electron motion occurs along two directions while in a nanowire it happens only in one direction.

*b. Effect of temperature*

Figure 2 shows the decay of the average spin vector at different temperatures for a nanowire and for a 2-D channel respectively at a driving electric field of 1 kV/cm. For a nanowire spin relaxation length increases from 210 nm at 300K to 575 nm at 150K, to 940 nm at 77K and 2.19 μm at 30K. For a 2D channel relaxation length increases from 12nm at 300K to 30nm at 150K and to 70nm at 77K. Thus spin relaxation length is observed to be a strong function of temperature and thus information remains preserved in the spin of electrons upto a greater distance.

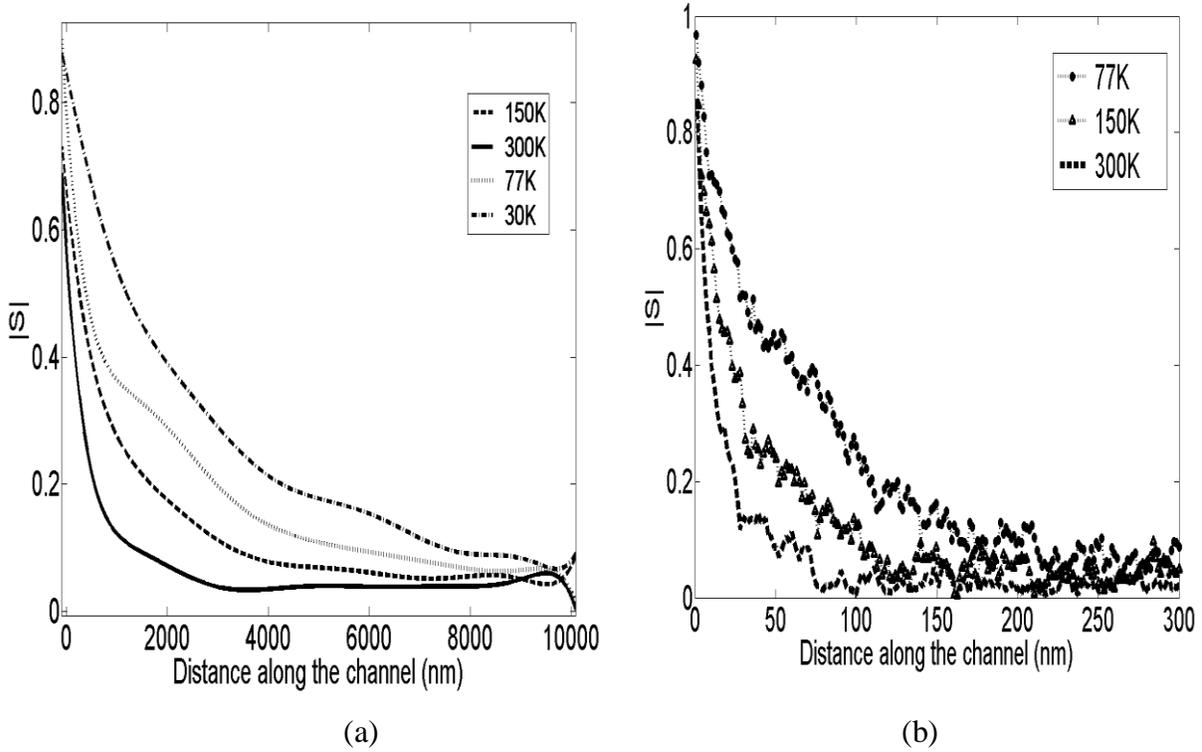

(a)    (b)

Fig.2: Variation of |S| with temperature. |S| along channel length at a driving electric field of 1kV/cm for (a) 1-D channel and (b) 2-D channel

On increasing the crystal temperature the phonon scattering rates increase. These increased scattering rates cause the electron to undergo scattering after very short time intervals and thus very short distances. This in turn randomizes the k vector rapidly and hence the precession vector also gets randomized within short distances from the source. Thus they get depolarized within smaller distances. This is again consistent with the fact that more random the motion of electron (due to increased temperatures) stronger is depolarization.

*c. Effect of applied electric field*

Figure 3 shows the decay of the magnitude of ensemble averaged spin vector at different driving electric fields for a nanowire and for a 2-D channel room temperature. The values of electric field used for analysis in our simulations are moderate enough to ensure that drift velocity saturation does not occur. The spin relaxation length for a nanowire changes from 193 nm at

100V/cm to 195 nm at 500V/cm, to 210 nm at 1 kV/cm, to 185 nm at 2kV/cm and to 222 nm at 5kV/cm. The spin relaxation length for a 2-D channel however shows a slight variation with applied electric field. The spin relaxation length for a 2-D channel first decreases from 16nm at 100V/cm to 13nm at 500V/cm and to 12nm at 1kV/cm. Thereafter it increases to 13nm at 2kV/cm and to 18nm at 5kV/cm. As compared to the dependence of spin dephasing on temperature, the dependence of spin dephasing on driving electric field is found to be weak and clearly nonmonotonic. This is because it is manipulated by two opposing factors- ensemble averaged drift velocity and scattering rates.

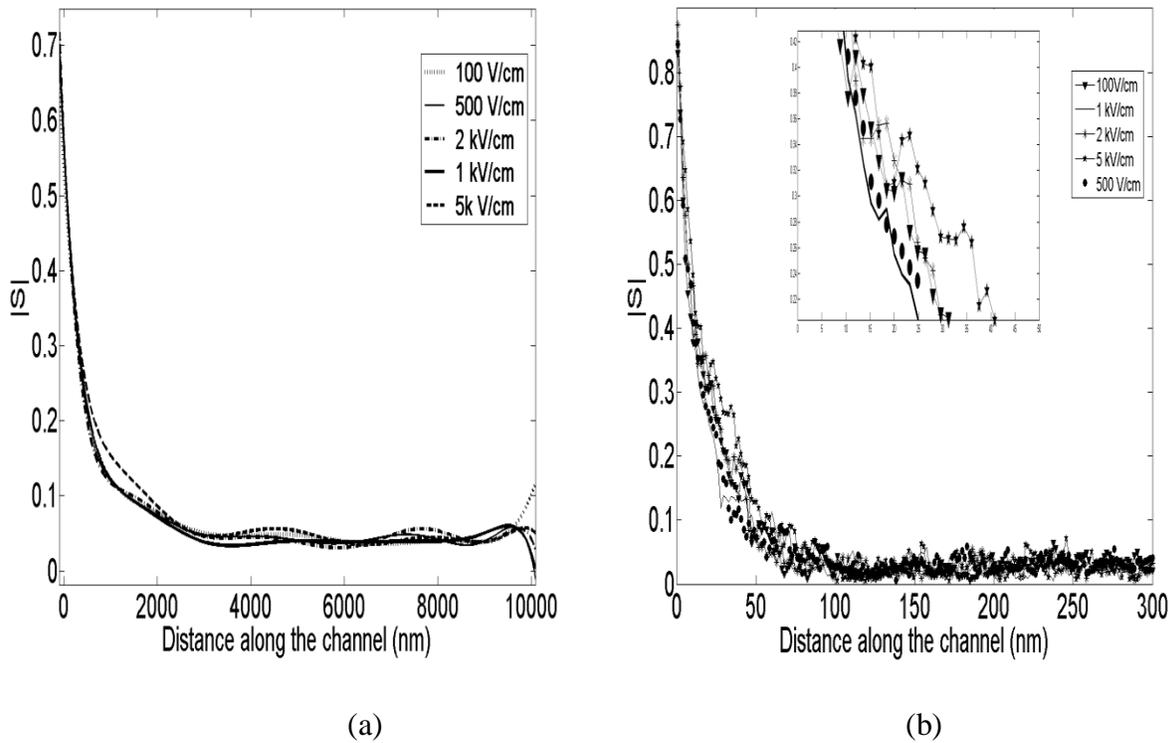

(a)            (b)

Fig.3: Variation of |S| with applied electric field. |S| along channel length at 300K for (a) 1-D channel and (b) 2-D channel

Any dominance of drift velocity over the scattering rates makes the electron and hence the spin penetrate further into the channel leading to larger spin relaxation lengths. The reverse happens when scattering rates dominate over drift velocity and the increased scattering rates dephase the spin faster. The overall effect is decided by the dominant effect amongst the two.

In our four subband model, the intersubband scattering saturates after a point as we increase the driving electric field. At higher driving electric fields the scattering rates remain fairly constant with only a slight variation. The drift velocity starts to dominate the scattering rates in this regime and spin relaxation length starts to increase. This explains the increase in relaxation length at higher electric fields.

## IV. CONCLUSION

In our work we show that confining the motion to only one direction can improve drastically upon the spin relaxation length (more than an order of magnitude to around 210 nm for a nanowire compared to 12nm for a 2-D channel at 300K and driving electric field of 1kV/cm). Thus the information stored in the spin of electrons remains polarized upto a larger length on using nanowires due to suppression of DP relaxation. This larger spin relaxation length leads us to believe that spintronic devices can be efficiently implemented with nanowires. Also we observe that the spin relaxation length in nanowires can be further increased by reducing the temperature, in which case it increases to 940 nm at 77K and 2.19 μm at 30K. Thus lowering the working temperatures can improve the performance of spintronic devices manifold.